\documentclass[12pt]{article}

\input{epsf}

\newcommand{\be}{\begin{equation}} 
\newcommand{\ee}{\end{equation}}

\begin{document}
\begin{titlepage}

\rightline{EFI-01-46} \rightline{hep-th/0110169}

\vskip 3cm
\centerline{ \Large \bf Three Dimensional de Sitter Gravity and
the Correspondence}

\vskip 2cm

\centerline{ \normalsize Bruno Carneiro da
Cunha\footnote{bcunha@theory.uchicago.edu}} 
\centerline{\sl Enrico Fermi Inst. and Dept. of Physics}
\centerline{\sl The University of Chicago}
\centerline{\sl 5640 S. Ellis Ave., Chicago IL 60637, USA}

\vskip 2cm

\begin{abstract}
Certain aspects of three dimensional asymptotically de Sitter spaces
are studied, with emphasis on the mapping between gravity observables
and the representation of the conformal symmetry of the boundary. In
particular, we show that non-real conformal weights for the boundary
theory correspond to space-times that have non-zero angular
momentum. Some miscellaneous results on the role of the holonomies and
isometry groups are also presented.  
\end{abstract}

\end{titlepage}

\newpage

\setcounter{footnote}{0}

\section{Introduction}

In the recent effort to understand quantum gravity in de Sitter (dS)
spaces \cite{bousso, strominger, witten, hull, balasubramanian,
klemm}, the role of  
asymptotic symmetries was particularly stressed. This is the group
which naturally acts on the conserved charges in classical theories of
gravitation \cite{wald1} and then on states of the Hilbert states of the
quantum theory. On the other hand, in cases where we perturb
space-time away from some simple vacuum, -- for instance, Minkowski, de
Sitter or the usual anti-de Sitter (AdS) space-times -- holography states
that basically all the information about the whole of space-time is
encoded in the conformal boundary. The asymptotic symmetry group would
then take for quantum gravity a similar role as the Poincar\'e group
has for the usual quantum field theory. 

The particular case of dS contrasts sharply with AdS in that there is
no model of a more fundamental description, no model derived from
string theory\footnote{despite some recent efforts in this
direction \cite{hull}.}. On the other hand, dS not only presents
itself as perhaps a more realistic background but also displays some
puzzles that have to be addressed before a background independent
formulation can even be dreamed of.

Among these is the finite observable entropy of dS spaces
\cite{bousso, bousso1, banks, stro-malda, kds}. The puzzle
relies on how a natural cut-off could be implemented that reduces the
degrees of freedom so drastically
to a finite number -- and perhaps worse, a finite number large enough to
be of any phenomenological interest. Another puzzle, perhaps related,
is the apparent instability of dS with respect to quantum
fluctuations \cite{ginsparg}. For instance, dS can ``decay'' by a
topology change process -- hence differently from the usual Minkowski
instability -- into many copies of itself. These questions have just
recently begun to be addressed, and they certainly possess some keys
to the understanding of quantum gravity in general.

Noteworthy progress in this direction was recently achieved by
Strominger \cite{strominger}, who proposed that the conjugacy between
scalar fields in the bulk and boundary operators worked much the
same way as in the AdS-CFT correspondence. Specifically, the conformal
weight of operators in the boundary was given by the asymptotic form
of the field profile as one approaches the conformal boundary -- in
the dS case, past and future infinity. A curious bound appears:
for fields whose mass was significantly larger than the de Sitter
mass, the corresponding operator would have complex
dimension. Although it poses no serious problem for the definition of
the boundary theory, which is after all Euclidean, the question of what
is the exact space-time interpretation of those operators still
remains. The clarification of this particular question in three
dimensions will be the aim of this short note.

The article is organized as follows: after a short introduction to the
Chern-Simons description of three dimensional gravity, emphasizing
the de Sitter case, we will present the general solution of three
dimensional de Sitter gravity and show that it can be seen as an
identification of global de Sitter under the action of a finite
group. We will close by discussing how these results can be used to
clarify some of the points raised above. 

\section{Chern-Simons and Asymptotic Data}

In three dimensions, the spin connection $\omega_{ab}$ is dual to a
vector field $\omega^a=\varepsilon^{abc}\omega_{ab}$. It was found in
\cite{csg} that the Einstein-Hilbert Lagrangian in three dimensions 
-- with a cosmological
constant term $\Lambda$ -- can be recast in terms
of the vector-valued one forms $A_{\pm}^a=\omega^a\pm
\sqrt{-\Lambda}e^a$, with $e^a$ being the triad. In the
$\Lambda<0$ case, both $A_{\pm}^a$ are real and independent. In the
$\Lambda>0$ case (which will be dwelled in from now on), they are the
complex conjugate of each other, and we will drop the subscript and
call $A_-$ by $A^*$. The action that results from the change of
variables has the Chern-Simons form\footnote{In \cite{stro-malda} or
\cite{banados} the reader will find a much more intelligible
discussion.}:

\be
S=\frac{1}{8 \pi G}\mbox{Im}\left[\int \mbox{Tr}(A\wedge d
A+\frac{2}{3}A\wedge A\wedge A)\right]
\ee
where $\mbox{Im}$ denotes the imaginary part. We will set $\Lambda=1$
and $4G=1$ from now on. The equations of motion
then simply state that $A$ is a locally flat connection:
\be
dA^a+{f^a}_{bc}A^a\wedge A^b=0
\ee
with ${f^a}_{bc}$ being the structure constants of the $SL(2,C)$
group. Being locally flat, the connection itself doesn't carry too
much information about the space-time. In parallel to the AdS case, one
has to provide some set of boundary conditions that introduce some
notion of ``asymptotic triviality'' and still allow for non-trivial
holonomies. From the fact that dS$_3$ has two disconnected conformal
boundaries at future and past infinity $t=\pm\infty$, one is naturally
led to the following boundary conditions:

\be
\left. ds^2\right|_{{\cal I}^-}=-dt^2+e^{-2t}dz
d\bar{z}+\ldots\label{asympast} 
\ee
and
\be
\left. ds^2\right|_{{\cal I}^+}=-dt^2+e^{2t}dz
d\bar{z}+\ldots\label{asymfuture} 
\ee

As we will see this turns out to be too restrictive. In a
holographic context, one argues that quantum gravity in dS consists of
a transfer matrix which relates the Hilbert spaces in the asymptotic
past and future. By requiring the two conditions above we are really
determining a particular state in both Hilbert spaces, and hence the
whole matrix element. In other words, if we restrict ourselves to the
gravity sector alone, the only space satisfying both of the
conditions above is global dS$_3$. We should then drop at least one of
the conditions. It became costumary to drop (\ref{asymfuture}), and
then the form of the connection can be derived much the same way as in
the AdS case \cite{coussaert}\footnote{The calculation was indeed
carried out for dS$_3$ recently \cite{klemm1}.}:
\be
A= -i \left(
\begin{array}{cc}
-\frac{1}{2}dt & e^{-t}dz \\
{\cal O}(e^t) & \frac{1}{2}dt
\end{array} \right) \label{bdry}.
\ee
where we represented $SL(2,C)$ with Pauli matrices. Following the
argumentation of section 4 in \cite{banados} one can see that the
general solution of the equations of motion with the boundary
conditions above is, up to a diffeomorphism:

\be
ds^2=-dt^2+(e^{-2t}+L(z)\bar{L}(\bar{z})e^{2t})dzd\bar{z}+L(z)dz^2+
\bar{L}(\bar{z})d\bar{z}^2 \label{solution}
\ee
with $L(z)$ an arbitrary holomorphic function and
$\bar{L}(\bar{z})=[L(z)]^*$. The actual proof is entirely analogous to
the AdS case (actually for AdS with Euclidean time, which is what is
done in \cite{banados}) and will not be shown here. The metric
(\ref{solution}) gives rise to the potential:
\be
A= -i \left(
\begin{array}{cc}
-\frac{1}{2}dt & e^{-t}dz \\
e^tL(z)dz & \frac{1}{2}dt
\end{array} \right) \label{potential}.
\ee

It is perhaps worth noting that the analogue of the Brown-Henneaux
symmetry is particularly clear in these coordinates, and not
particularly surprisingly so. Setting
$\bar{L}(\bar{z})=0$ for the sake of clarity, one sees that the
transformation:
\be
\delta t=\frac{1}{2}\partial \epsilon,\;\;\;\delta
z=\epsilon,\;\;\;\delta \bar{z}=\frac{1}{2}e^{2t}\partial^2\epsilon
\label{transf}
\ee
is a symmetry if we allow $L(z)$ to transform as:
\be
\delta L=-(\epsilon \partial L+2\partial\epsilon
L+\frac{1}{2}\partial^3\epsilon ).
\ee
From which we can rederive the central charge \cite{strominger,
nojiri, myung} by recovering the dependence on $l$ and $G$. The point
of view naturally taken is this case is that the 
transformation (\ref{transf}) is a diffeomorphism that changes the
asymptotic charges and then is not a gauge transformation but rather
a global transformation.

So now we turn to the problem of relating the function $L(z)$ to
physical quantities of the space-time metric such as mass and angular
momentum. 

\section{Cartography}

We will now turn to the relation of the metric (\ref{solution}) with
the three-dimensional Kerr-de Sitter solution \cite{kds}:
\be
ds^2=-\left(M-r^2+\frac{J^2}{4r^2}\right)d\bar{t}^2+\frac{dr^2}
{\left(M-r^2+\frac{J^2}{4r^2}\right)}+r^2\left(d\varphi-
\frac{J}{2r^2}d\bar{t}\right)^2 \label{kds}.
\ee
For that consider (\ref{solution}) with constant $L(z)=L_0$. The
mapping 
\be
t'=t-\frac{1}{4}\ln ( L_0\bar{L}_0),\;\;x =
\sqrt{L_0}z+\sqrt{\bar{L}_0}\bar{z},\;\;iy =
\sqrt{L_0}z-\sqrt{\bar{L}_0}\bar{z}
\ee
turns the metric into
\be
ds^2=-dt'^2+\cosh^2t'dx^2+\sinh^2t'dy^2 \label{solution2}.
\ee
The Kerr-de Sitter metric can also be turned into (\ref{solution2}) by
the following change of coordinates:
\be
\sinh t' = \sqrt{\frac{r^2-r_+^2}{r_+^2+r_-^2}},\;\;
y = r_+~\bar{t}+r_-~\varphi,\;\;
x = r_-~\bar{t}-r_+~\varphi. \label{changekds}
\ee
where $M=r_+^2-r_-^2$ and $J=2r_+r_-$. Note that only the region
$r>r_+$, $t'>0$, and thus lying outside the cosmological horizon can
be mapped. In this coordinate change we find
\be
L_0=\left(r_+-ir_-\right)^2=
M-iJ\equiv\lambda^2 
\ee
and also that $r_-~y-r_+~x$ is
identified with $2\pi (r_+^2+r_-^2)$ translations. For
(\ref{solution}) that means $z+\bar{z}=\varphi\in [0,2\pi]$. Note
that $L_0$ is complex for 
non-zero $J$. Spatial slices have then the
topology of a cylinder, whose non-contractible loop is indeed
necessary for the existence of a non-trivial holonomy for $A$. Also,
it fits in the usual description of mass sources as conical
defects in global dS where the spatial slices are spheres.

It is also interesting to study the global structure of the metric
(\ref{solution}), or, alternatively, (\ref{solution2}). We will
restrict ourselves to the case of zero angular momentum, in which $x$
is an angular variable with a conical defect $x\in
[0,2\pi\lambda]$ but 
$y$ is not restricted. By defining:
\be
\tan \hat{t} = \sinh t' \cosh y,\;\; \cos \theta =\frac{\sinh t' \tanh
y}{\sqrt{ \cosh^2t'-\tanh^2y}},\;\; \phi=x,
\ee
one turns (\ref{solution2}) into the familiar form of global
dS \cite{hawking}:
\be
ds^2=\sec^2 \hat{t}(-d\hat{t}^2+d\theta^2+\sin^2\theta d\phi^2)
\ee
The patch covered by $\{t', x, y\}$ can be seen in Figure 1. Note
that it covers both ${\cal I}^+$ and ${\cal I}^-$ which also happen to
be the holographic screens of dS.

It may not be clear that the null boundary of the hourglass is at
$y\rightarrow \infty$. To see that one just needs to consider the
inverse transformation:
\be
\tanh y=\frac{\cos\theta}{\sin\hat{t}}
\ee
which defines the boundary for $\sin\hat{t}=\pm\cos\theta$. Also, the
point at $\hat{t}=0$ is merely a coordinate singularity, just like in
the AdS version of (\ref{solution}). 

We now turn to the study of the observables and the identifications
that introduce non-trivial holonomies from global dS$_3$.

\begin{figure}[htb]
\begin{center}
\epsfxsize=6cm \epsfbox{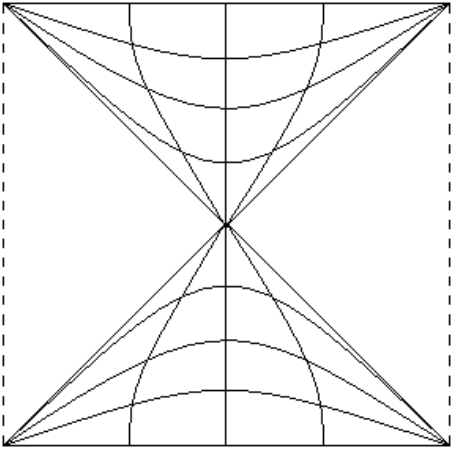}
\caption{{\it The patch of de Sitter space covered by the coordinates
(\ref{solution}) for real $L_0$ is the central hourglass-like
region. The horizontal lines and vertical lines correspond to $t'$ and
$y$ being constant respectively. The null boundary is at
$y=\pm\infty$ in (\ref{solution2}).}}
\end{center}
\end{figure}

\section{Holonomies and Identifications}

Since $A$ is a flat connection, it can be written locally by $A=i{\cal
G}^{-1} d 
{\cal G}$, where ${\cal G}$ is an element of $SL(2,C)$. For $A$ as
in (\ref{potential}), ${\cal G}$ can be written as:

\be
{\cal G}= \left(
\begin{array}{cc}
\Psi_1'(z) & \Psi_1(z) \\
\Psi_2'(z) & \Psi_2(z)
\end{array} \right)
\left(
\begin{array}{cc}
e^{t/2} & 0 \\
0 & e^{-t/2} 
\end{array} \right)
\ee
with $\Psi_{1,2}(z)$ the solutions of the Schr\"odinger equation
$(\partial^2_z - L(z))\Psi(z)=0$, set to have the Wronskian
equal to one.

For $L(z)=L_0=\lambda^2$ constant it is useful to define a $2\times 2$
coefficient matrix $B$ as: 
\be
g(z)=
\left(
\begin{array}{cc}
\Psi_1'(z) & \Psi_1(z) \\
\Psi_2'(z) & \Psi_2(z)
\end{array} \right)
\equiv
B
\left(
\begin{array}{cc}
\lambda e^{\lambda z} &  e^{\lambda z} \\
-\lambda  e^{-\lambda z} &  e^{-\lambda z}
\end{array} \right)
\ee
Note that the actual elements of $B$ are not so important, since they
can be gauged away by a global $SL(2,C)$ transformation, but
we have to require $\mbox{det} B =\frac{1}{2\lambda}$. Using the
fact that the $z$ 
plane has the topology of a cylinder, the monodromy can be easily
computed:
\be
g(z+\pi)=B e^{2\pi \lambda \sigma_3}B^{-1}g(z),
\ee
and then the holonomy is $W=2\cosh (\pi \lambda)$.

Finally, we will show that (\ref{solution}) can be obtained from
global dS$_3$ via a proper identification. One way of accomplish that
is by embedding the hourglass region (\ref{solution2}) in $SL(2,C)$:
\be
g=e^{i (x+iy) \sigma_2}e^{2t\sigma_3}e^{i (x-iy)\sigma_2}
\ee
and the action of the $SL(2,C)$ group is given by $g\rightarrow h g
h^*$. The particular group element that implements the isometry 
that brings $x+iy$ to $x+iy+2\pi\lambda$ is 
\be
h=e^{2\pi i\sigma_2\lambda},
\ee 
and then the space (\ref{solution}) can be though of as
the quotient of global dS$_3$ under the action of the group generated
by $h$.

\section{Discussion}

In this short note we have collected some results about asymptotic
dS$_3$ spaces which may help clarify the role of holography in the
quantum theory with that background. Specifically, we have mapped the
conformal weights of operators of the boundary theory into space-time
quantities like the mass and the angular momentum of the space. It was
shown that the weights are complex in general, in accordance with the
results of \cite{strominger}.

Given the form of the general metric solution (\ref{kds}), one might
be tempted to work in cosmological patch where time is a Killing
vector field. The point of view taken here is different in which we
stress the role of the asymptotic symmetries. This is in consonance
with holographic arguments which infer that the states of quantum
gravity in de Sitter can be encoded in the conformal boundary of
space-time. It is particularly clear in this language the space-time
role of the Virasoro generators of the boundary conformal field
theory.

Of course, there are a few caveats. It is unclear how exactly the
description given in terms of the asymptotic data relates to what a
given observer sees. Recovering the observer's point of view may be
more than a technicality given that Susskind {\it et al.} argued on
generic grounds \cite{susskind} that a quantum description of gravity
based on local degrees of freedom cannot see beyond a horizon. Being
more specific, the local formulations of observables for each side of
the horizon are not compatible with one another. Also, one would like
to verify which mechanism in the boundary conformal theory is
responsible for the finiteness of the dS states. One expects that the
correlation between different states in the boundary theory to be
large, thus reducing its entropy. Note that here the same scenario
envisioned by \cite{cunha} happens: there are patches of space-time
which have alternative projections on either past or future infinity
screens. Following that argument, the language best suited for coping
with these issues, especially if one's goal is to describe local
states in the bulk of space, may be that of a correlated or entangled
state (see also \cite{maldacena}.) 

As this work was in its final states of preparation, we learned of a
similar work \cite{bbm} which has overlaps with the results listed
here. After this work was made publicly available we became aware of
the work by Park \cite{park} in which he also derives the space-time
interpretation of the asymptotic conformal theory. Unlike that work,
we argue, based on the causal structure of de Sitter, that the
boundary conformal theory is defined on a pair of two-spheres instead of a
cylinder. We also gave a detailed account on how the Virasoro symmetry 
acts on the space-time metric, and discussed the role of
identifications and holonomies. We have also found a real central
charge for the Virasoro symmetry, corroborating the results of
\cite{strominger, klemm, banados1, nojiri, myung, bbm}.

\section*{Acknowledgements}

I would like to thank Emil Martinec for helpful discussions and
suggestions on the manuscript. I am also indebted to Ben Craps, Will
McElgin, Vasilis Niarchos and Li-Sheng Tseng who in their patience
helped me put my thoughts in order.

\end{document}